\documentclass[journal]{IEEEtran}  %

\usepackage{multirow}     %
\usepackage{booktabs}     %
\usepackage{amsmath,amssymb,amsfonts}
\usepackage{rotating}
\usepackage{graphicx}
\usepackage{algorithm}
\usepackage{algorithmic}
\usepackage{adjustbox}
\usepackage[graphicx]{realboxes}

\usepackage[numbers,sort&compress]{natbib}
\usepackage{float}
\usepackage{textcomp}
\usepackage{xcolor}
\usepackage{amsmath}
\usepackage{caption}
\usepackage{subcaption}
\usepackage{xcolor}
\usepackage{pgfplots} \pgfplotsset{compat=1.18}
\usepackage{pgfplots}
\usepackage{tikz}
\usepackage{pgfplotstable}
\usetikzlibrary{pgfplots.groupplots}

  {\endgroup}%

\begin{document}

\bstctlcite{IEEEexample:BSTcontrol}
\title{A 129FPS Full HD Real-Time Accelerator for 3D Gaussian Splatting}

\author{\IEEEauthorblockN{Fang-Chi Chang, and Tian-Sheuan Chang, \textit{Senior Member, IEEE}}
\thanks{This work was supported by the National Science and Technology Council, Taiwan, under Grant 111-2622-8-A49-018-SB, 110-2221-E-A49-148-MY3, 113-2221-E-A49-078-MY3, and 113-2640-E-A49-005. The authors are affiliated with the Institute of Electronics, National Yang Ming Chiao Tung University, Taiwan. (e-mail: p311580042.11@nycu.edu.tw, tschang@nycu.edu.tw) This paper has been accepted to be published in IEEE Transactions on Visualization and Computer Graphics. }%
\thanks{Manuscript received XXXX XX, 2025; revised XXXX XX, XXXX.}
}
\maketitle

\begin{abstract}

Rendering large-scale, unbounded scenes on AR/VR-class devices is constrained by the computation, bandwidth, and storage cost of 3D Gaussian Splatting (3DGS). We propose a low-power, low-cost 3DGS hardware accelerator that renders full-HD images in real time, together with a hardware-friendly compression pipeline that combines iterative Gaussian pruning and fine-tuning, progressive spherical harmonics (SH) degree reduction, and vector quantization of all SH coefficients and colors. The scheme achieves a $51.6\times$ model-size reduction with  a 0.743 dB PSNR loss. The accelerator uses a frame-level pipeline that integrates point-based culling and projection with tile-based sorting and rasterization, skips zero-Jacobian matrix multiplications (reducing processing elements by 63\% and computation by 53\%), and adopts comparison-free tile-based sorting with deterministic latency. Implemented in a TSMC 28-nm process at 800 MHz, the design occupies $0.66~\text{mm}^2$ with 1.1438 M gates and 120 kB SRAM, consumes 0.219 W, and delivers 1219 Mpixels/J at 267.5 Mpixels/s, enabling 1080p at 129 FPS. Overall, it is $5.98\times$ smaller in area, $5.94\times$ higher throughput, and delivers $7.5\times$ higher energy efficiency than prior 3DGS accelerators.

\end{abstract}

\begin{IEEEkeywords}
3D Gaussian Splatting, hardware accelerators, model compression, real-time rendering
\end{IEEEkeywords}

\section{Introduction}
\label{chapter:introduction}

Recently, 3D Gaussian Splatting (3DGS) has emerged~\cite{3DGS} as an popular novel view synthesis method, using a set of 3D Gaussians to represent scenes, enabling high-quality real-time rendering on GPUs. However, 3DGS models typically require larger memory footprints, making them difficult to deploy on resource-constrained portable devices used in augmented reality (AR) and virtual reality (VR) applications. Thus, hardware acceleration coupled with effective model compression techniques becomes critical.

A large body of work have been proposed to reduce the storage and bandwidth footprint of 3DGS through attribute compression, structural reparameterization, and pruning~\cite{morgenstern2024compact, compact3DGS, lightgaussian, papantonakis2024reducing, fang2024mini, mallick2024taming, zoomers2025progs, radl2024stopthepop}. 
However, most of the software acceleration methods are targeted for GPU-based execution, which is not hardware-friendly. Beyond software acceleration on GPUs, hardware accelerators have also attracted attention in recent years~\cite{gscore, LeeGS, gsnorm}. However, these designs incur high area and power costs because they implement the original 3DGS algorithm, and they cannot meet the low-power and high-throughput requirements of real-time full-HD rendering for AR/VR edge devices~\cite{overbeck2018system}.

To overcome these limitations, this paper introduces a dedicated hardware accelerator for real-time rendering of 3DGS models on portable AR/VR devices, targeting full-HD resolution at 120 FPS per eye~\cite{overbeck2018system}. First, for model compression, we modify the lightweight model, LightGaussian~\cite{lightgaussian}, to higher compression ratios (51.6$\times$ vs 15$\times$) by iterative Gaussian pruning and SH degree reduction, and by applying VQ to all SH coefficients and colors.
Second, our hardware architecture proposes a frame-level pipeline with mixed granularity for high hardware utilization and low latency: point-based processing in preprocessing stages and tile-based processing during rendering. This design further reduces complexity by skipping zero-Jacobian matrix multiplications in the projection path and adopts tile-based, comparison-free sorting with $\langle\text{tile-id},\text{depth}\rangle$ keys to match front-to-back $\alpha$-blending at fixed per-output latency. 
The final hardware implementation achieves full-HD rendering at 129 FPS, which is 3$\times$ higher throughput and 5.98$\times$ smaller area when compared to the prior 3DGS accelerator~\cite{gscore}.

The remainder of the paper is organized as follows. Section~II shows the related work. Section III details the software-level compression methods used for 3DGS. Section~IV describes the proposed hardware accelerator design. Experimental results are provided in Section~V, followed by concluding remarks in Section~VI.

\section{Related Work}
\label{sec:related_work}

The rapid adoption of 3D Gaussian Splatting (3DGS) has driven research to reduce storage, rendering-time computation, and latency for interactive and streaming scenarios. Existing efforts mainly trade off reducing bytes per Gaussian (e.g., coding, quantization, and codebooks/VQ) versus reducing Gaussian count (e.g., pruning/compaction and SH-degree reduction). The survey in~\cite{bagdasarian20253dgs} summarizes these trade-offs and their impact on representation size and runtime costs such as sorting and blending.

Many works compress attributes while largely preserving Gaussian count. Morgenstern et al.~\cite{morgenstern2024compact} reorganize attributes into 2D grids for codec efficiency. EAGLES~\cite{eagles} uses an encoding--decoding scheme for memory reduction. Compact3DGS~\cite{compact3DGS} applies VQ to color and SH parameters, while LightGaussian~\cite{lightgaussian} reduces SH degree and quantizes low-salience components. ContextGS~\cite{wang2024contextgs} introduces an anchor-level context model, HAC~\cite{chen2024hac} incorporates hash-grid-assisted context, and CodecGS~\cite{lee2025compression} compress 3DGS using optimized feature planes with standard video codecs. Related budgeting methods also include resolution-aware pruning and SH allocation~\cite{papantonakis2024reducing}.

Complementary to byte reduction, compaction and pruning reduce Gaussian count to lower compute, memory traffic, and sorting complexity. Mini-Splatting~\cite{fang2024mini} combines densification and simplification with constrained SH to ease queuing and sorting, and TAMING 3DGS~\cite{mallick2024taming} uses budget-constrained training to produce predictable model sizes suitable for fixed on-chip buffering.

Progressive and streaming-oriented methods improve time-to-first-paint and perceived latency. PRoGS~\cite{zoomers2025progs} prioritizes high-contribution splats under bandwidth limits, aligning with termination strategies that stop blending when remaining contributions are negligible. StopThePop~\cite{radl2024stopthepop} reduces popping via hierarchical resorting; hardware implementations often prefer tile-local approximations to avoid global synchronization and excess traffic. Training-time optimizations can also improve compactness and inference behavior; for example, Kheradmand et al.~\cite{kheradmand20243d} formulate training as MCMC sampling to enhance Gaussian relocation.

Several works propose hardware accelerators for 3DGS and related pipelines. GSCore~\cite{gscore} introduces shape-aware tests and hierarchical sorting. Lee et al.~\cite{LeeGS} study group sorting, reconfigurable matrices, and importance-aware SH handling. GSNorm~\cite{gsnorm} uses normalization and LUT-quantized rendering to decouple dependencies. Earlier tile-reordered heaps~\cite{weyrich2007hardware} also relate to coherence-friendly ordering, although modern 3DGS places stricter bandwidth and buffering demands.

Our work adopts a hardware--compression co-design for resource-constrained devices. We tailor compression to hardware constraints and design the accelerator around tile-level ordering and buffering under limited on-chip SRAM. Unlike accelerators that assume unmodified models or rely on complex global ordering and preprocessing, the proposed co-design achieves higher throughput and energy efficiency with lower area cost.

\section{Proposed Algorithm}
\label{chapter:3D Gaussian Splatting Model Compression}

\subsection{Background: 3D Gaussian Splatting (3DGS)}
\label{sec:bg_3dgs}
\subsubsection{Pipeline Overview}
Novel view synthesis (radiance field rendering) generates photorealistic images of a scene from its underlying 3D representation.
3DGS models the scene using a set of Gaussian ellipsoids, enabling fast rendering by rasterizing these Gaussians into high-quality images. The inference process in 3DGS consists of four main steps: near-plane culling, projection, Gaussian sorting, and rasterization.

\begin{itemize}

  \item \textbf{Near-plane culling:}
    The process begins with a lightweight near-plane visibility test that discards Gaussians behind the camera near plane to avoid unnecessary projection and rendering computations.

    \item \textbf{Projection:}
    After visibility filtering, the surviving Gaussians, along with view parameters, are projected from 3D space onto the 2D image plane.
    The features of the projected 2D Gaussian points, such as depth, 2D coordinates, opacity, radius, and color, are computed based on the splatting algorithm.

    \item \textbf{Gaussian sorting:}
    Next, the Gaussian points are sorted by depth to determine their order along the view ray, which is essential for accurate color blending and visibility in the final image.

    \item \textbf{Rasterization:}
    Finally, the 2D image is divided into tiles (e.g., 16$\times$16 pixels).
    Using the computed Gaussian features, pixel colors within each tile are calculated simultaneously.
    The final image is then assembled by merging the rasterized tiles.
\end{itemize}

Although 3DGS maps well to modern desktop GPUs and achieves real-time rendering, its memory footprint and bandwidth demands make the original algorithm poorly suited to resource-constrained devices such as mobile SoCs and standalone AR/VR headsets.

\subsubsection{Projection to Screen Space and the Jacobian}
\label{sec:bg_3dgs_jacobian}
Assuming a pinhole camera model with intrinsics $(f_x,f_y,c_x,c_y)$, a 3D point $\mathbf{x}=[X,Y,Z]^\top$ in camera coordinates is projected to the image plane by
\begin{equation}
u = f_x \frac{X}{Z} + c_x,\quad
v = f_y \frac{Y}{Z} + c_y .
\label{eq:pinhole_projection}
\end{equation}
The Jacobian of the projection $\pi(\mathbf{x})=[u,v]^\top$ with respect to $\mathbf{x}$ is
\begin{equation}
J(\mathbf{x})=\frac{\partial (u,v)}{\partial (X,Y,Z)}=
\begin{bmatrix}
\frac{f_x}{Z} & 0 & -\frac{f_x X}{Z^2}\\
0 & \frac{f_y}{Z} & -\frac{f_y Y}{Z^2}
\end{bmatrix}.
\label{eq:projection_jacobian}
\end{equation}
A first-order approximation propagates the 3D Gaussian covariance $\Sigma_{3D}$ into screen space via
\begin{equation}
\Sigma_{2D} \approx J(\mathbf{x})\,\Sigma_{3D}\,J(\mathbf{x})^\top .
\label{eq:cov_propagation}
\end{equation}
In practice, $\Sigma_{2D}$ is converted to a 2D elliptical footprint (often represented as a conic) for rasterization.
Importantly, $J(\mathbf{x})$ contains identically zero terms (Eq.~\eqref{eq:projection_jacobian}), which enables hardware to skip multiplications by zero without changing numerical results.

\subsubsection{Depth-Sorted Alpha Compositing and Early Termination}
\label{sec:bg_3dgs_alpha}
For each pixel, 3DGS performs depth-sorted front-to-back alpha compositing over the Gaussians that cover the pixel.
Let $\alpha_i\in[0,1]$ denote the effective opacity contribution of the $i$-th Gaussian at the pixel (after evaluating the Gaussian footprint), and let $\mathbf{c}_i$ be its view-dependent color.
The accumulated color $\mathbf{C}$ and transmittance $T$ are updated as
\begin{align}
\mathbf{C}_{i} &= \mathbf{C}_{i-1} + T_{i-1}\,\alpha_i\,\mathbf{c}_i, \label{eq:alpha_color_update}\\
T_{i} &= T_{i-1}(1-\alpha_i), \label{eq:alpha_transmittance}
\end{align}
with initialization $\mathbf{C}_0=\mathbf{0}$ and $T_0=1$.
Early termination stops processing additional Gaussians once the remaining transmittance becomes sufficiently small, e.g.,
\begin{equation}
T_i < \tau,
\label{eq:early_termination}
\end{equation}
because subsequent contributions are bounded by $T_i$ and have negligible impact on the final color.

\subsection{Analysis and Overview of the Proposed Approaches}
We adopt the original 3D Gaussian Splatting (3DGS) renderer~\cite{3DGS} as our algorithmic baseline. The 3DGS algorithm is not hardware-friendly and faces four main design challenges, as follows. 

\subsubsection{Substantial Memory Usage}
Typically, unbounded 360-degree scenes require millions of Gaussian points, leading to storage requirements in the gigabyte range. To mitigate this, we propose a model compression scheme based on LightGaussian~\cite{lightgaussian}.

\subsubsection{Accessing and Computing Unnecessary Gaussian Points}
During the inference phase, once the viewpoint and direction are determined, our analysis shows that approximately 42\% of Gaussian points are considered unnecessary via near-plane culling. 
    This percentage is view-dependent; for example, when zooming out so that the entire scene fits inside the view frustum, nearly all Gaussians can become visible and the culling benefit decreases.
In the rendering process, during \(\alpha\)-blending, only about 50\% of Gaussian points contribute to the pixel color. Once the accumulated transmittance falls below a threshold, more than half of the points are terminated early, making their sorting a waste of computational resources. To address this, we adopt near-plane culling~\cite{3DGS} to eliminate parameter access for invisible areas, skip zero-Jacobian matrix multiplications, and apply early termination in \(\alpha\)-blending~\cite{3DGS, gscore}.

\subsubsection{Sorting Millions of Gaussian Points}
In the rendering process, 3DGS requires depth sorting of all Gaussian points. However, sorting the entire list of Gaussian points is extremely resource-intensive and becomes a bottleneck in the rendering pipeline for real-world scenes. To solve this, we adopt tile-based, comparison-free sorting, restricting sorting within a tile, which requires only linear delay cycles and low hardware complexity.

\subsubsection{High Variability in the Number of Gaussian Points per Scene, Tile, and Pixel}
The number of Gaussian points per scene and per view within the same scene has high variability, resulting in an unpredictable number of Gaussian points required per tile and per pixel. This situation complicates pipeline management, parallelism, and memory control in hardware design to achieve a general solution. To address this, we propose using a key–value global prefetch buffer to reduce latency and improve local memory utilization.

For the above challenges, Among these challenges, the first one is addressed by the proposed model compression, while the other three are addressed through hardware design optimizations.

\subsection{Proposed Model Compression}
\begin{figure}[tbp]
\centering
\includegraphics[width=1.0\linewidth]{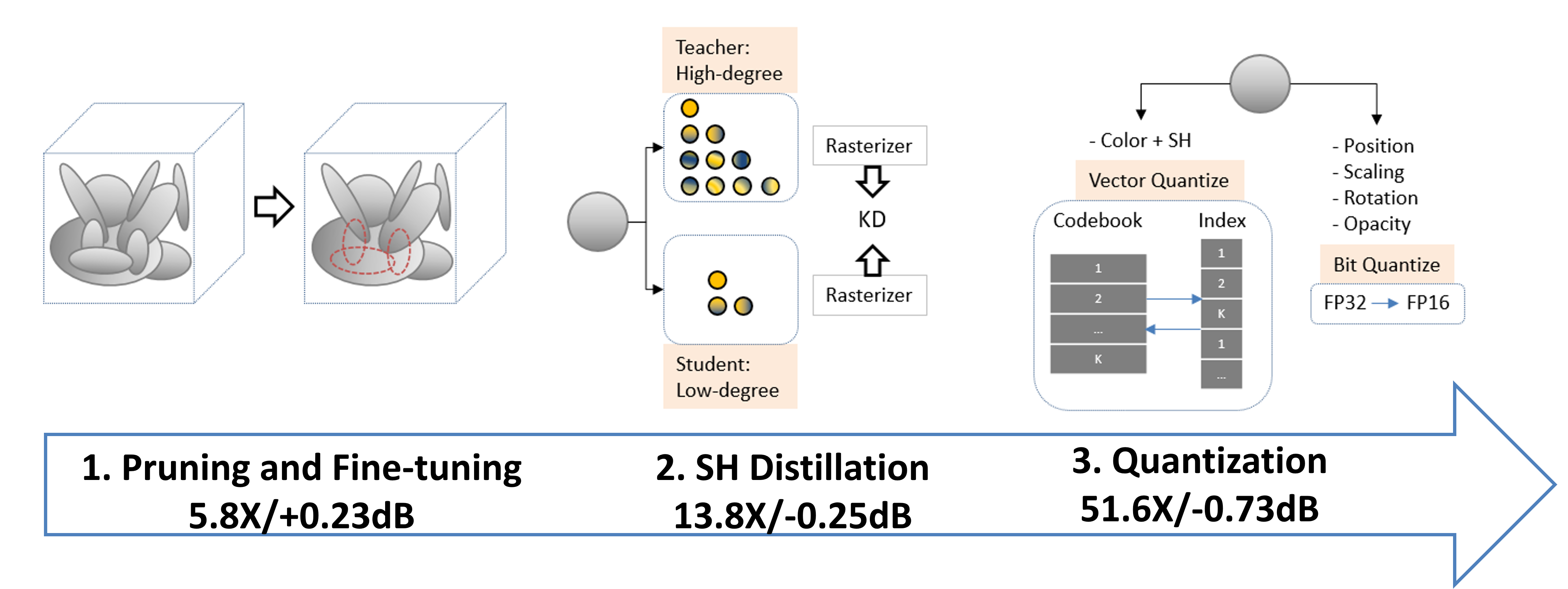}
\caption{Overview of our model compression method.}
\label{3_5_software_overview}
\end{figure}

Model compression reduces computational complexity and memory usage, and is therefore essential for hardware-oriented 3DGS deployment. The compression pipeline used in this work, shown in Fig.~\ref{3_5_software_overview}, is derived from LightGaussian~\cite{lightgaussian}, which combines Gaussian pruning and fine-tuning, SH distillation, VQ for SH coefficients and colors, and FP16 quantization. Although this pipeline provides a strong baseline, its compression ratio is insufficient to meet our design target, namely full-HD rendering at 120~FPS per eye under a sub-0.5~W power budget. Moreover, directly increasing the compression ratio of LightGaussian leads to noticeable PSNR degradation and visible detail loss. We therefore adapt the pipeline for edge deployment, where Gaussian count and bytes per Gaussian point jointly determine workload and power. Our goal is to achieve a predictable, hardware-feasible design under strict on-chip buffering constraints. In particular, we make the following modifications:

\begin{itemize}
    \item \textbf{Iterative pruning with intermediate fine-tuning:}
    Instead of applying aggressive pruning in a single step, we prune the model progressively and insert fine-tuning between pruning stages. This improves recovery of visual quality while reducing Gaussian count to the level required by the hardware design.

    \item \textbf{Progressive SH-degree reduction via iterative distillation:}
    Instead of directly lowering the SH degree in one step, we distill the representation progressively. This reduces bytes per Gaussian point with a smoother quality--compression tradeoff and avoids the abrupt quality drop caused by one-shot reduction.

    \item \textbf{VQ applied to all SH coefficients and colors:}
    We apply vector quantization to all SH coefficients and colors, rather than only low-salience subsets, to further reduce per-Gaussian storage at a bounded quality cost.
\end{itemize}

These modifications jointly reduce Gaussian count and per-Gaussian storage, thereby lowering workload and data movement in the hardware pipeline. Compared with LightGaussian, the resulting pipeline achieves a substantially higher compression ratio (51.6$\times$ vs.\ 15$\times$) at a bounded quality cost. Fig.~\ref{3_5_software_overview} summarizes the average model-size reduction and quality metrics of each compression step across datasets, and detailed results are given in Section~V.

\section{Hardware Implementation}
This section first presents the hardware challenges and our proposed solutions, and then describes the detailed hardware design.

\subsection{Challenges and Proposed Solutions}
\subsubsection{Reduction of Unnecessary Gaussian Points}

\paragraph{Near-plane culling}
\label{sec:near_plane_culling}
We adopt the view near-plane culling method with axis-aligned bounding boxes (AABB) from the original 3DGS algorithm, which checks depth to determine whether a point should be removed.

Let $z$ denote the Gaussian center depth in camera coordinates and let $\Delta z \ge 0$ be the half-extent of the AABB along the camera $z$-axis.
We define the depth interval as $[z_{\min}, z_{\max}] = [z-\Delta z,\; z+\Delta z]$ and cull the Gaussian if
\begin{equation}
z_{\max} < z_{\text{near}},
\label{eq:near_plane_cull}
\end{equation}
where $z_{\text{near}}$ is the camera near-plane depth.
This is a near-plane-only test (not a full six-plane frustum test), chosen to maximize pruning benefit per hardware cost.

\begin{figure}[tbp]
\centering
\includegraphics[width=0.8\linewidth]{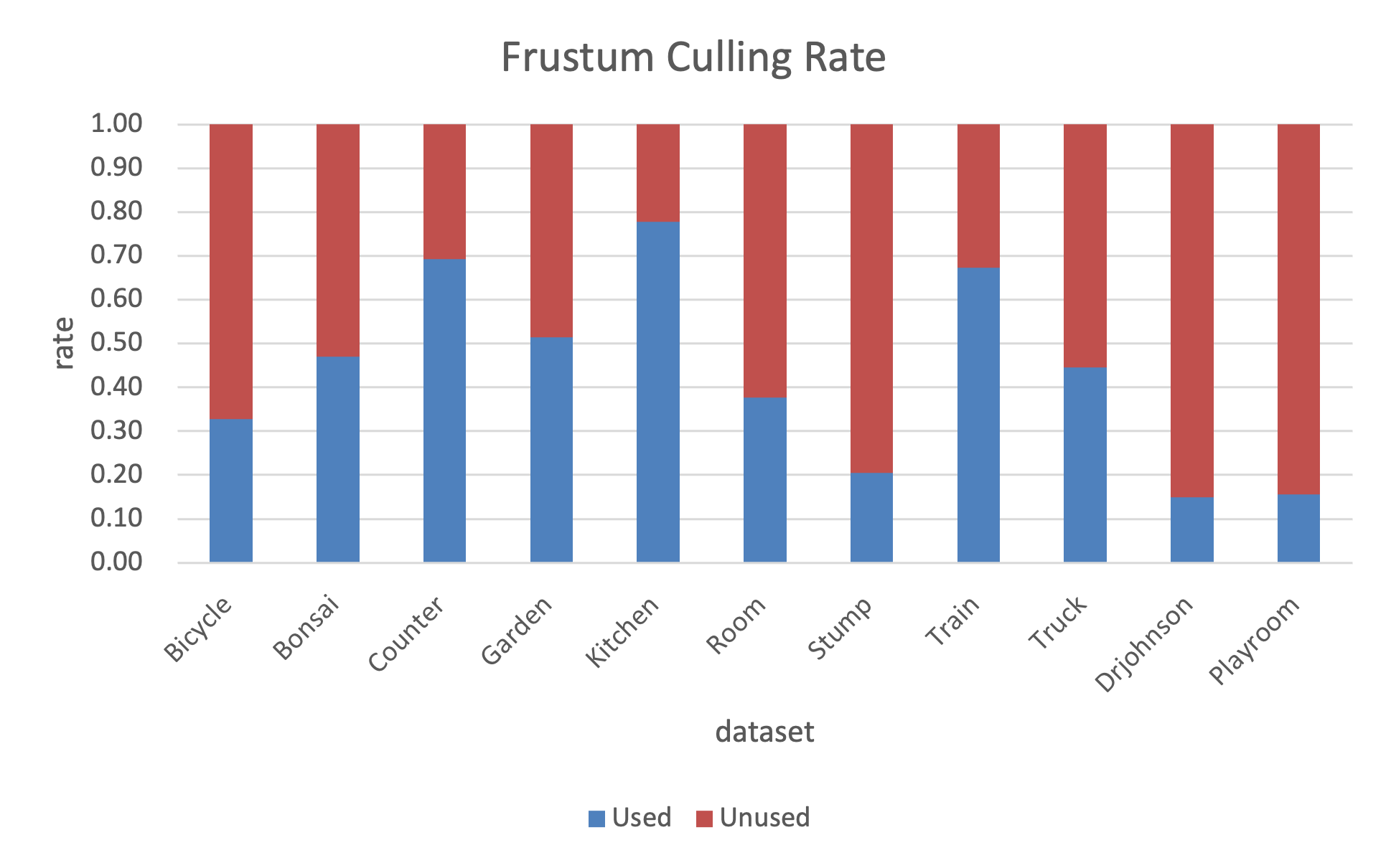}
\caption{Analysis of the near-plane culling rate for the compressed model.}
\label{4_2_frustum_cullnig_usedandunused}
\end{figure}

Applying near-plane culling to our compressed model yields a 56\% culling rate (Fig.~\ref{4_2_frustum_cullnig_usedandunused}), meaning that 56\% of unnecessary Gaussian point computations and memory accesses are saved. In comparison, the near-plane culling rate of the uncompressed 3DGS model is approximately 60\%. Our average near-plane culling rate is slightly lower because we apply 51.6$\times$ compression beforehand.

\paragraph{Skip Zero-Jacobian Matrix Multiplication}
In the projection stage, the Jacobian matrix contains several zero elements  (see Sec.~\ref{sec:bg_3dgs_jacobian} and Eq.~\eqref{eq:projection_jacobian}). We can skip the corresponding multiplications, reducing computation by 52.53\% compared to the projection step in the original 3DGS algorithm. Table~\ref{Table_zero-skipping} summarizes the operation counts in the projection stage. This optimization enables us to reduce the PE array from a 4$\times$4 configuration to a 6$\times$1 configuration, saving 63\% of the PEs.

\begin{table}[tbp]
\centering
\caption{Comparison of the improvement by skipping zero Jacobian matrix multiplication.}
\label{Table_zero-skipping}
\begin{tabular}{l|c|cccl}
                  & Total & +  & $\times$   & / & - \\ \hline
w/o zero-skipping & 198   & 78 & 112 & 7 & 1 \\
w/ zero-skipping  & 94    & 46 & 42  & 5 & 1 \\
Reduction         & 104   & 30 & 70  & 2 & 0
\end{tabular}
\end{table}

\begin{figure}[tbp]
\centering
\includegraphics[width=0.8\linewidth]{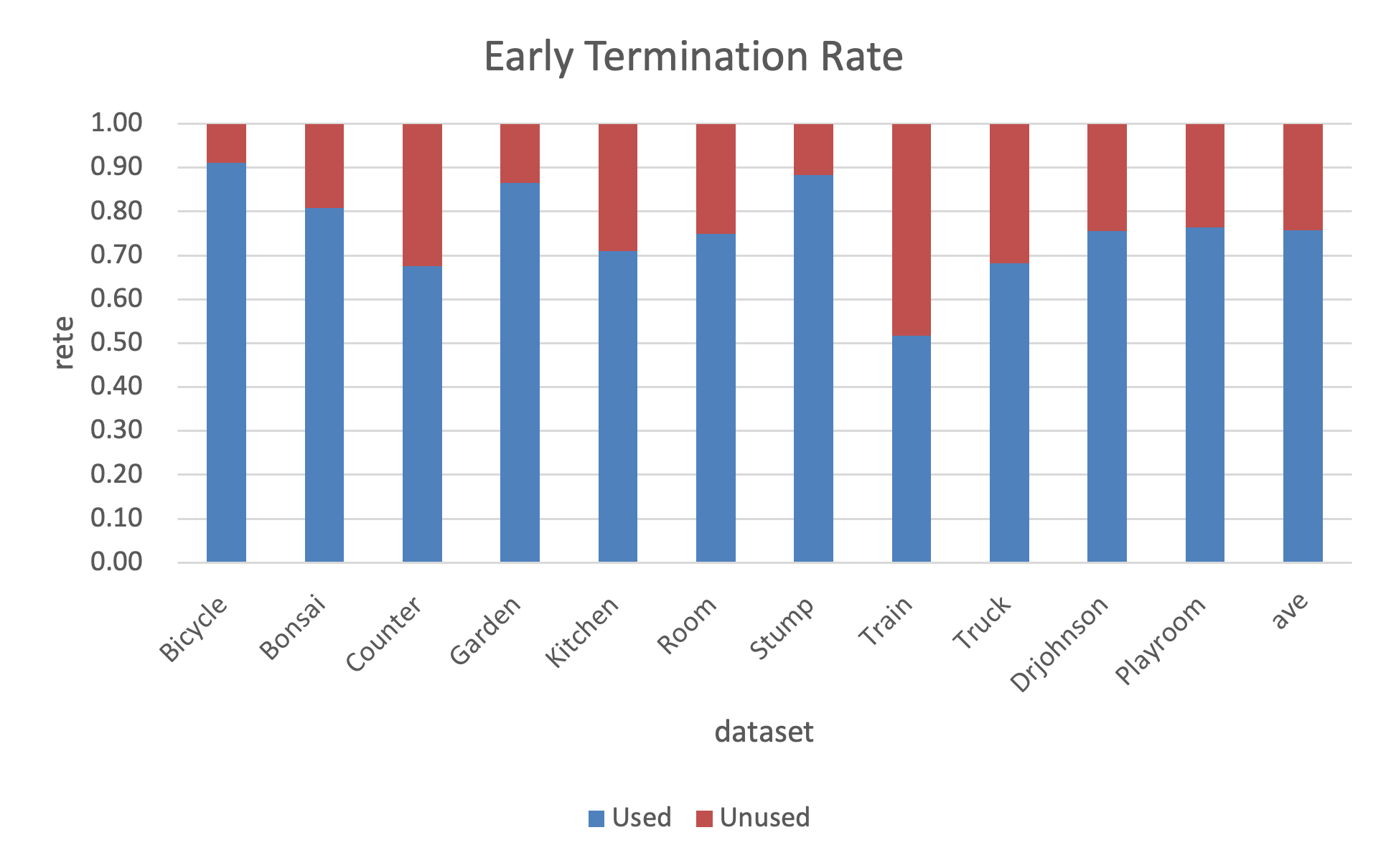}
\caption{Analysis of early termination rate.}
\label{4_4_early_termination_rate}
\end{figure}

\paragraph{\(\alpha\) Pruning and Early Termination}
We also adopt the \(\alpha\) pruning and early termination from 3DGS~\cite{3DGS} and GSCore~\cite{gscore} for lower complexity and memory access. Fig.~\ref{4_4_early_termination_rate} shows the number of unnecessary Gaussian points excluded by the early termination mechanism  (see Sec.~\ref{sec:bg_3dgs_alpha} and Eq.~\eqref{eq:alpha_color_update}--\eqref{eq:early_termination}). In the original uncompressed 3DGS model, approximately 50\% of Gaussian points are unused due to early termination. In our proposed compressed 3DGS model, the average ratio of unused Gaussian points due to early termination is reduced to 24.3\%. This reduction indicates that our approach can save around 24\% of computational resources in the subsequent volume rendering and sorting stages.

\subsubsection{Sorting Millions of Gaussian Points}
\label{4-subsection:Sorting Millions of Gaussian Points}

In the original 3DGS pipeline, sorting millions of Gaussian points for alpha-blending relies on global radix sort on GPUs, achieving linear time complexity but incurring high memory overhead from histograms, prefix sums, and scatters. This leads to increased gate counts due to parallel scans and variable latency, making it unsuitable for low-power AR/VR hardware and predictable hardware scheduling.

We adopt tile-based, comparison-free sorting based on the parallel comparison-free algorithm~\cite{parallel-comparison-free-sorter, comparison-free-sorter} but with several proposed optimizations to enhance its efficiency. The comparison-free sorting algorithm~\cite{parallel-comparison-free-sorter} offers several notable features and advantages, particularly in hardware implementations, such as being comparator-free, having simple logic gates for O(N) deterministic cycles, enabling max detection in one cycle, parallelism, and scalability.

Their algorithm sorts $N$ $n$-bit elements in $\mathcal{O}(N)$ time using concurrent and sequential phases, guided by the Element Vector Table (EVT). The EVT is initialized with 1s, tracks unsorted elements and updates each cycle to exclude the largest element (LE). In the concurrent phase, $k$ parallel clusters process partitions of the $N$ $n$-bit elements, starting from the most significant bit, progressively filtering out smaller values while retaining the larger ones through bitwise AND operations and multiplexer-based selection within cascaded blocks. The outputs from all clusters are then passed to the sequential phase, where a linear array of $k$ cascaded blocks further refines the selection, ultimately identifying the LE. This process repeats $N$ times until the EVT contains all 0’s, signaling that all $N$ $n$-bit elements have been sorted.

To further optimize the algorithm~\cite{parallel-comparison-free-sorter}, we propose the following methods:

\textbf{Sign Bit Skipping:}  
In the near-plane culling stage, points with a depth of less than 0 are excluded. Thus, the sign bit is always 0. With this, the operation on the sign bit can be removed to reduce unnecessary computations.

\textbf{Data Conflict Resolution:}  
When performing the largest element detection, it is necessary to handle duplicate values. If the returned EVT contains more than one "1", it could lead to errors in the final result. To simplify the implementation of sorting datasets that contain duplicate values, we adopt a straightforward approach involving bitwise inverter, bitwise AND, and an adder.
\begin{equation}\label{eq6}
    Fo_{NEW} = Fo \ \& \ ( \sim Fo+1'b1)
\end{equation}

\subsection{Proposed Architecture}
\label{4-section:Proposed Architecture}

\begin{figure}[tbp]
\centering
\includegraphics[width=1.0\linewidth]{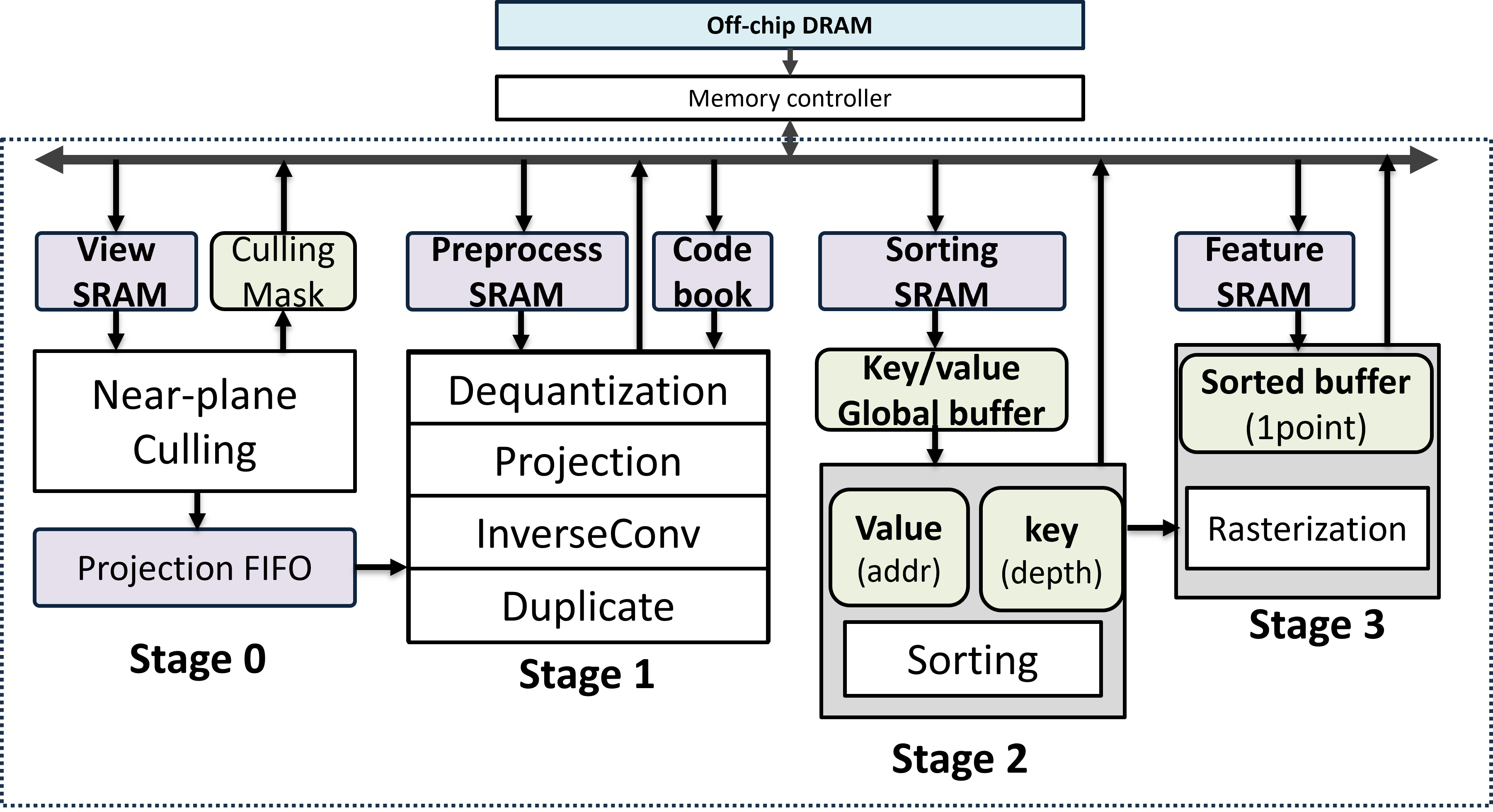}
\caption{Proposed system architecture.}
\label{4_1_Hardware_Arch}
\end{figure}

\begin{figure}[tbp]
\centering
\includegraphics[width=1.0\linewidth]{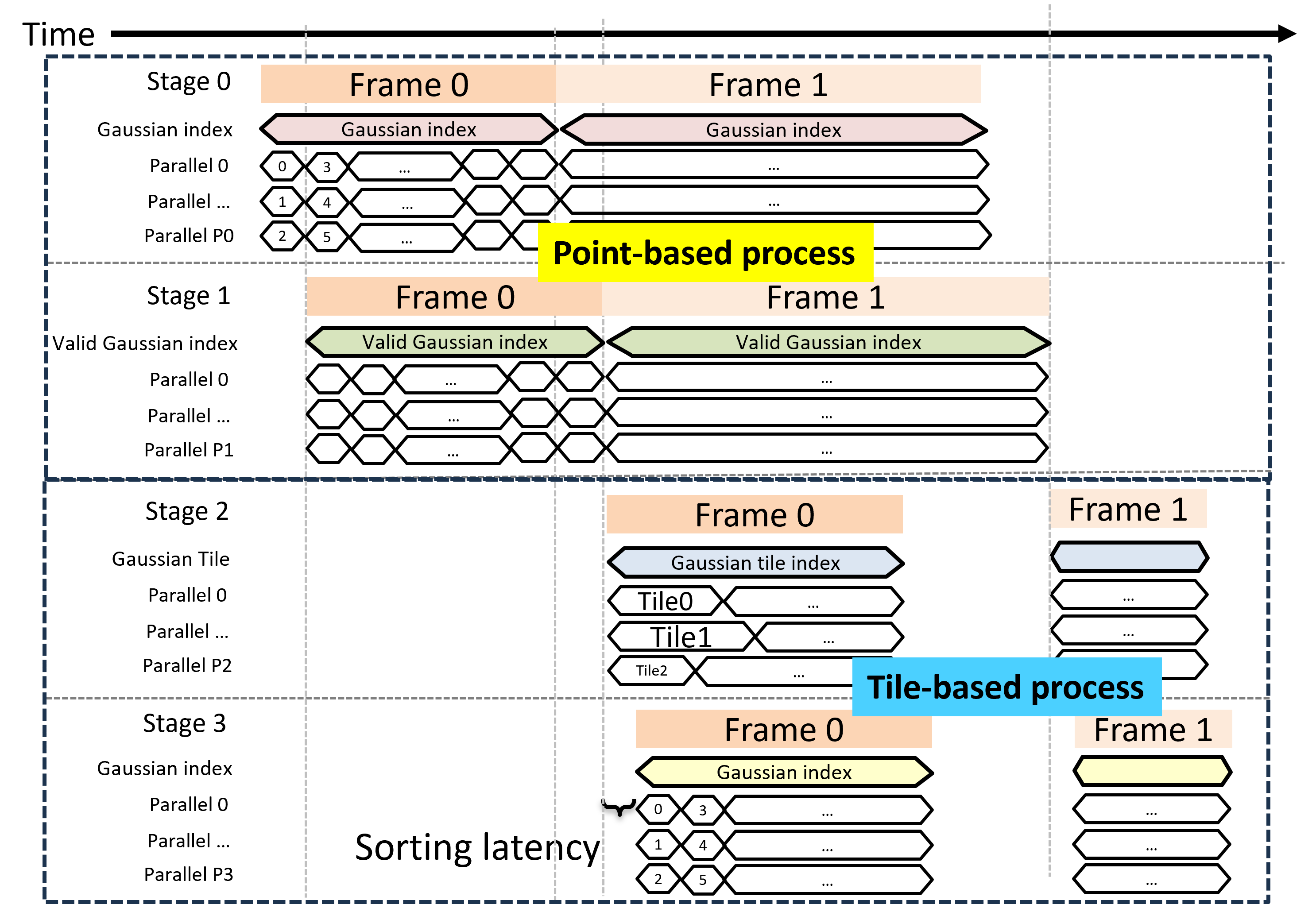}
\caption{Timing diagram of the proposed design. $N$ refers to the number of Gaussian points, which may vary depending on the scene or view.}
\label{4_15_timing_diagram}
\end{figure}

Fig.~\ref{4_1_Hardware_Arch} shows the hardware architecture with four pipeline stages. The first two stages comprise the preprocessing step, which contains the near-plane culling unit to eliminate parameter access for invisible regions in Stage 0, and the projection unit to project 3D Gaussians onto the 2D image plane with skipping zero-Jacobian matrix multiplication in Stage 1 
 (see Sec.~\ref{sec:bg_3dgs_jacobian} and Eq.~\eqref{eq:projection_jacobian}). The last two stages comprise the rendering step, with the Gaussian sorting unit in Stage 2 and the rasterization unit in Stage 3. In this pipeline, we isolate the near-plane culling unit into a dedicated stage because it must process the largest number of Gaussian points. With this isolation, we can easily optimize the computational resources.

Fig.~\ref{4_15_timing_diagram} shows the timing diagram of the proposed pipeline. In particular, the preprocessing step employs point-based processing since it processes the Gaussian points. In contrast, the rendering step utilizes tile-based processing since it processes the image tiles. Due to the data dependency between the preprocessing and the rendering steps, the sorting operation in Stage 2 must wait until all points have been projected onto the 2D plane before execution can begin. This ensures that the global sorting can be completed accurately and efficiently, as it requires the full set of projected data to maintain the correct order and coherence across the entire dataset. Therefore, we chose to implement a frame-level pipelining design between these two steps.

\subsubsection{Preprocessing Step}
\paragraph{Stage 0: near-plane culling}
\label{4-subsection:Stage 0}
\begin{figure}[tbp]
\centering
\includegraphics[width=1.0\linewidth]{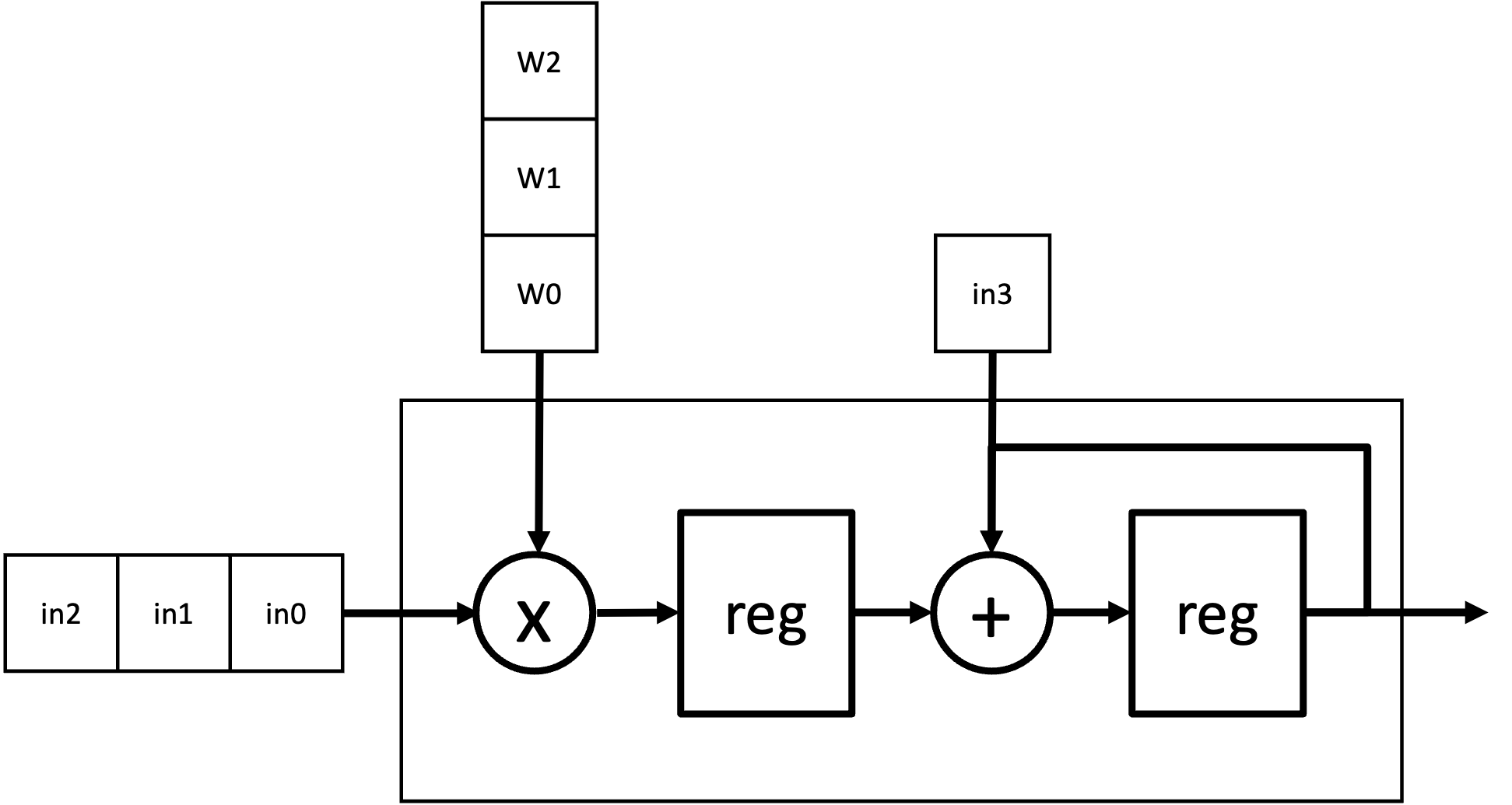}
\caption{ The proposed near-plane culling unit with one row of the view matrix in[3:0] and (x, y, depth) (w[2:0] in the figure). }
\label{4_3_stage0_module}
\end{figure}
Fig.~\ref{4_3_stage0_module} illustrates the design of the proposed near-plane culling unit, which implements the AABB depth-interval near-plane test in Eq.~\eqref{eq:near_plane_cull}. With the proposed culling optimization to reduce complexity and memory access, this design operates with only a single MAC (Multiply-Accumulate) unit and has a latency of 4 cycles, ensuring that it performs efficiently while maintaining low resource consumption. Because the computations are performed at 800 MHz in FP16 format, a D flip-flop (DFF) is inserted between each operation to ensure proper timing and data stability.

\paragraph{Stage 1: Projection}
\label{4-subsection:Stage 1}

\begin{figure}[tbp]
\centering
\includegraphics[width=0.5\linewidth]{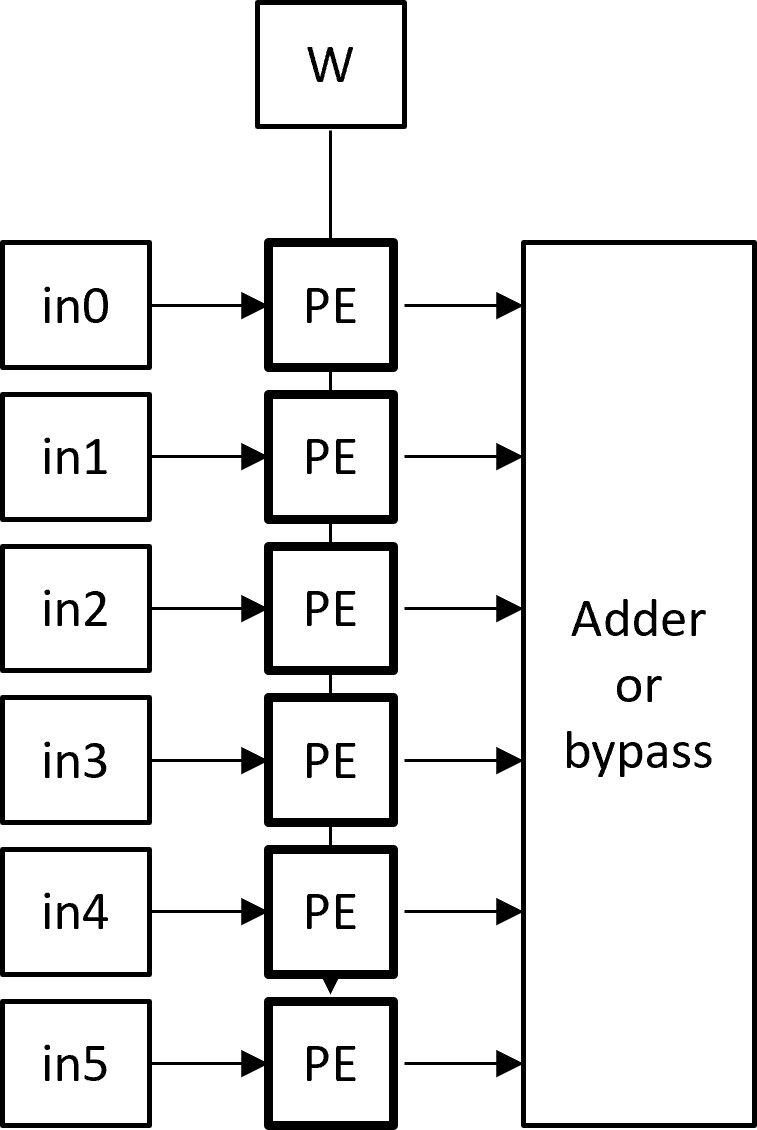}
\caption{The computation datapath of the Stage 1. }
\label{fig:stage1}
\end{figure}
From a hardware-systems perspective, Stage~1 is throughput-critical because it must process all visible Gaussians each frame  and has significant data dependencies. 

To address this, we partition Stage 1 into smaller submodules based on these data dependencies, including the SH, projection, and inverse covariance computation submodules. Since all these submodules involve only matrix multiplications, we can share the computational resources between submodules. Thus, based on the complexity requirement, we use a 6$\times$1 one-dimensional multiplier-accumulator array in Fig.~\ref{fig:stage1}  to compute all these matrix multiplications and sum or directly output the result with the adder/bypass unit. 
This design uses weight broadcasting to maximize data reuse. The hardware utilization of this array is 51.6\%. This relatively low utilization is primarily due to the high data dependency inherent in the processes handled by this array. The data dependencies limit the ability to fully parallelize operations, as certain computations must wait for the completion of others, thereby reducing the overall efficiency of the PEs.

\subsubsection{Rendering Step}
The rendering step is tile-based, with a tile size set to 16$\times$16 pixels, while the preprocessing step is point-based. 
Each 16×16 tile is assigned to an independent SRAM bank with a fixed entry depth and address offset controller. The index is used to look up the corresponding value stored in the codebook.
The sorting algorithm used in Stage 2 is particularly notable for its ability to output the current maximum value in a linear sorting delay. This feature aligns well with the \(\alpha\)-blending process in the rasterization, where one point is calculated at a time. This approach not only increases processing speed but also reduces the memory required for rasterization.

\begin{figure}[tbp]
\centering
\includegraphics[width=1.0\linewidth]{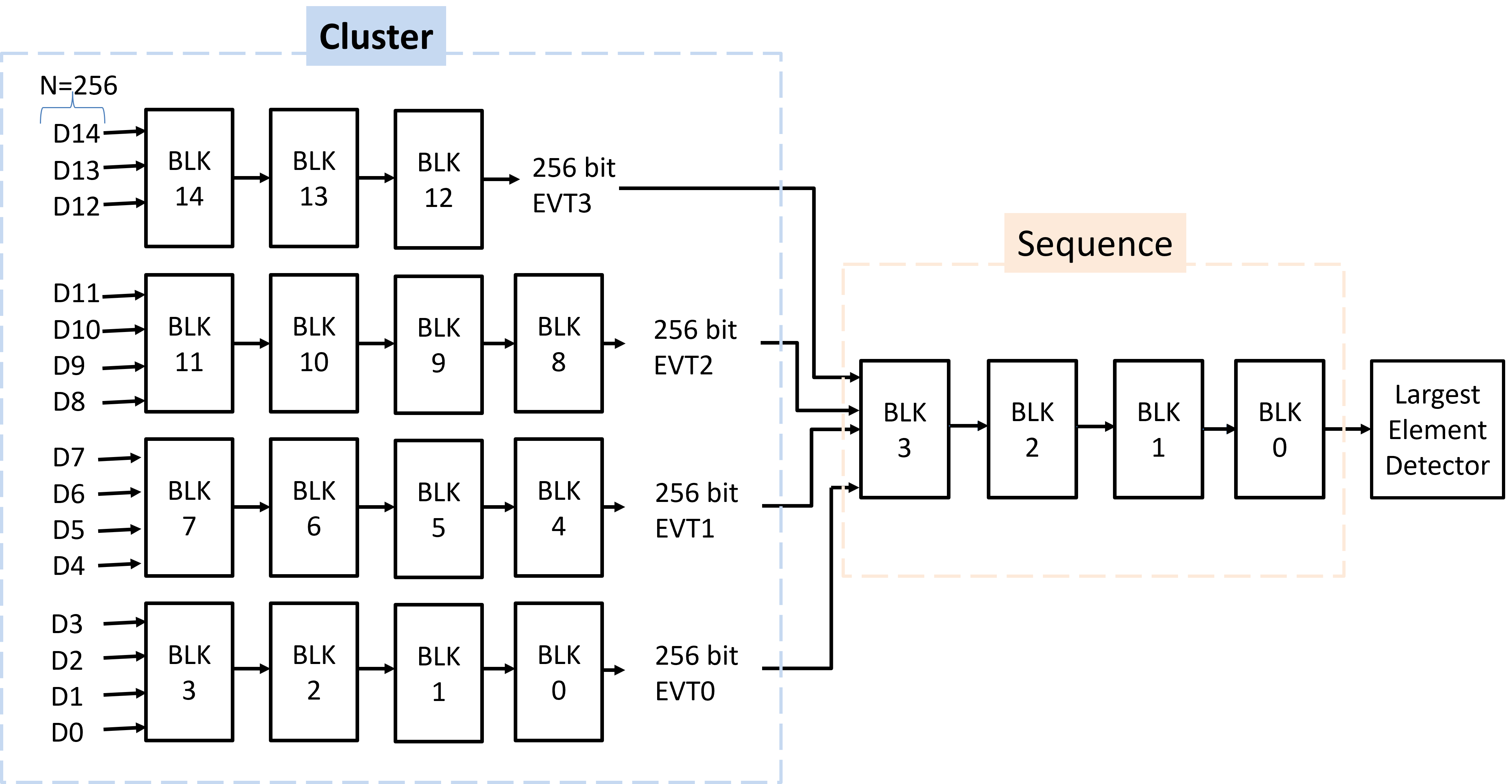}
\caption{Proposed Gaussian sorting unit.}
\label{4_8_sort_256x15}
\end{figure}

\paragraph{Stage 2: Gaussian Sorting}
\label{4-subsection:Stage 2}
Fig.~\ref{4_8_sort_256x15} shows the proposed Gaussian-sorting module with four-way sub-sorter parallelism, where the sub-sorter design is based on the comparison-free sorter~\cite{parallel-comparison-free-sorter, comparison-free-sorter}. Each sub-sorter processes 256 data elements at a time. This design choice is influenced by the fact that the critical path of the sub-sorter lies in the delay of the $N$-input OR gate, where $N$ represents the number of data elements being sorted. The experimental results indicated that a sub-sorter designed to process 512 data elements is more suitable for operation at frequencies below 500\,MHz in a 28\,nm process environment, which does not meet our target frequency of 800\,MHz. This is the main reason why we selected the specific value for $N$ in the sub-sorter design. This module can output the current maximum value every 2 cycles. This module will sort 256$\times$15-bit data. The sign bit is skipped since all numbers are positive. The 15-bit data is divided into (3, 4, 4, 4) bit groups, sent to the cluster stage for sorting, and then processed by the sequence stage to get the address of the largest element.

\begin{figure}[tbp]
\centering
\includegraphics[width=0.6\linewidth]{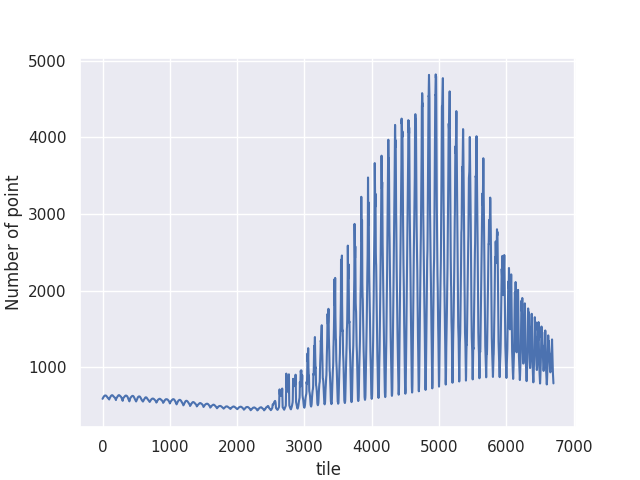}
\caption{Tile density analysis on the "Bicycle" dataset.}
\label{4_11_bicycle_tile_density}
\end{figure}

In this design, a major scheduling problem for the pipeline is the high variability of the Gaussian points, as shown in Fig.~\ref{4_11_bicycle_tile_density}. This raises the problem of deciding the hardware parallelism and buffer size for better hardware utilization. In the figure, most tiles contain around 1000 Gaussian points. However, there is significant variability, ranging from a few hundred to 5000. Additionally, there is considerable variation even between adjacent tiles. For such high variability, one intuitive approach is to set an upper bound on the processed Gaussian points per tile to ease hardware design. However, discarding those points beyond the bound will result in image "holes" that cannot be easily compensated.

To adapt to the high variability, this design adds a shared key–value global buffer for four sub-sorters. Each sub-sorter includes a 4\,KB key buffer and a 4\,KB value buffer to process 2000 Gaussian points per tile. The excess points are stored in the 12\,KB key and 12\,KB value global buffers to reduce area cost and maximize hardware utilization.

Our Stage-2 uses a comparison-free key sorter tailored to 3DGS: each splat emits a ⟨tile-id, depth⟩ key, and tiles are consumed front-to-back by \(\alpha\)-blending. This yields fixed per-output latency and high sustained throughput, while avoiding O(log n) priority-queue operations and matching the one-key-at-a-time consumption within each tile. By contrast, the work in \cite{weyrich2007hardware} employs a constant-throughput pipelined heap that reorders splats mainly by tile index to improve cache coherence and impose a cyclic tile scan. Their design overlaps insert/delete across heap levels (a modified SIFT-DOWN), effectively achieving constant-time operations in a pipeline with extra complexity. The two approaches are complementary: the pipelined-heap is a general, cache-friendly reordering Gaussians that can reduce external bandwidth by trading some on-chip cache for reordering, whereas our sorter is a specialized, low-control-overhead Gaussians tuned to the 3DGS key structure and the consumption pattern of \(\alpha\)-compositing.

\begin{figure}[tbp]
\centering
\includegraphics[width=1.0\linewidth]{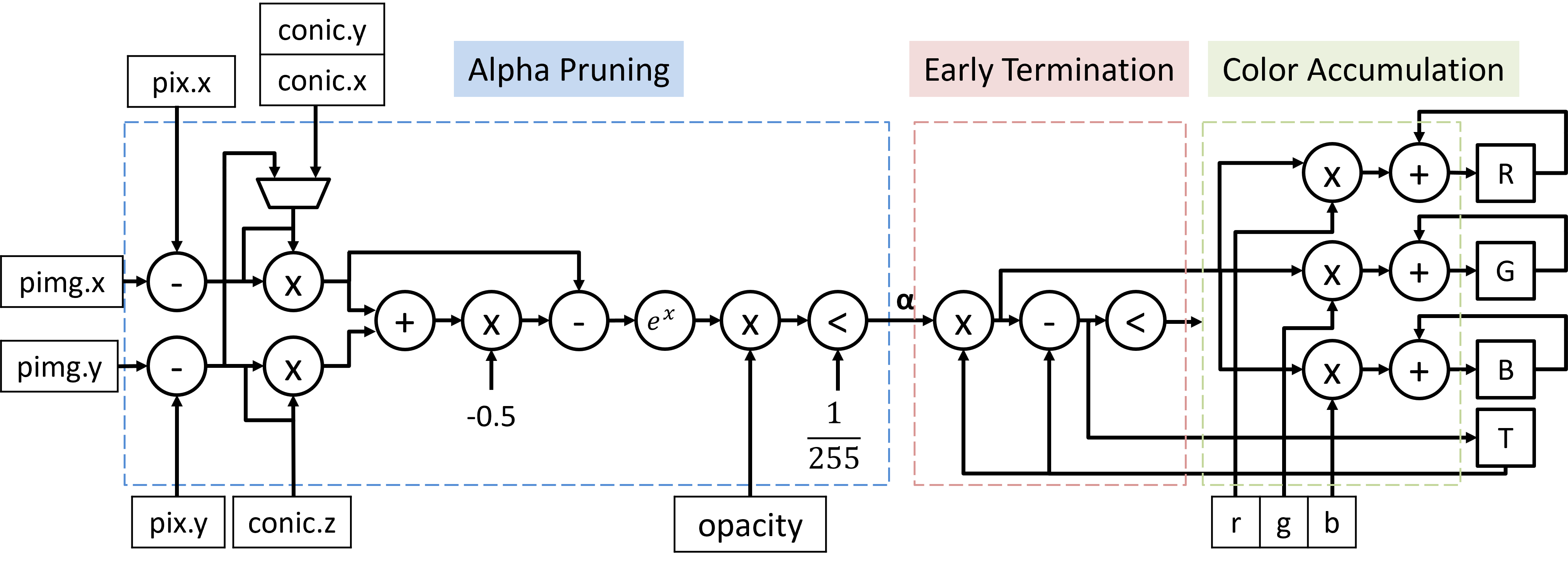}
\caption{Proposed rasterization stage.}
\label{4_7_reander_module}
\end{figure}

\paragraph{Stage 3: Rasterization}
\label{4-subsection:Stage 3}
Fig.~\ref{4_7_reander_module} illustrates the detailed architecture of our proposed rasterization module, which is subdivided into three main stages: \(\alpha\)-pruning, early termination, and color accumulation. The \(\alpha\)-pruning stage prunes unnecessary Gaussian points based on their \(\alpha\) values, thus optimizing the rendering process by reducing the number of points that proceed to the next stages. The early termination stage determines whether the accumulated pixel color has reached a saturation point, beyond which further processing would not significantly alter the final output (see Eq.~\eqref{eq:alpha_color_update}--\eqref{eq:early_termination}). This stage performs multiplication and subtraction operations followed by a comparison, allowing for the early termination of the rendering process for certain Gaussian points, thereby saving computational resources. The final color accumulation stage accumulates the color contributions (R, G, B channels) from the Gaussian points that pass the previous stages. The operations are performed separately for each color channel and the transmittance ($T$)(see Eq.~\eqref{eq:alpha_transmittance}). The results are then combined to produce the final rendered image.

\section{Experimental Results and Evaluation}
\label{chapter:Experimental Result}

\subsection{Experimental Settings}
\label{5-section:Experimental Settings}
\subsubsection{Datasets and Metrics}
\label{5-subsection:Datasets and metrics}
The datasets for evaluation include Mip-NeRF360~\cite{mipnerf360} for complex, unbounded environments, Tanks\&Temples~\cite{tandt} for highly detailed and realistic scenes, and Deep Blending~\cite{deepblending} for smooth blending of image patches and seamless integration in multiview image synthesis. With diverse scenarios, these datasets demonstrate the effectiveness of our approach.

\subsubsection{Implementation Details}
\label{5-subsection:Implementation details}
The 3DGS model~\cite{3DGS} is adopted as our pretrained model and comparison baseline. For our pruning strategy, each iterative pruning step lasts for 5000 steps, 
    during which we fine-tune the pruned model using a pure image-space L1 loss between the rendered and ground-truth RGB images, instead of the combined L1 + D-SSIM loss used in the original 3DGS training.
The initial learning rates match those in 3DGS: position = 0.00016, opacity = 0.05, scaling = 0.005, and rotation = 0.001. For the SH distillation process, the iteration count is set to 4000. The mean square error (MSE) is used for the VQ process.

\begin{table}[btp]
\centering
\caption{Analysis of gate count, area, and power.}
\setlength{\tabcolsep}{2pt}
\begin{tabular}{lcccccc}
\hline
               & \multicolumn{2}{c}{Gate Count (k)} & \multicolumn{2}{c}{Area ($\mu$m$^2$)} & \multicolumn{2}{c}{Power (mW)} \\ \hline
Stage 0        & 12.9               & 1.1\%        & 4887                      & 0.70\%              & 2.14              & 1.00\%    \\
Stage 1        & 71.6               & 6.3\%        & 27069                     & 4.10\%              & 6.66              & 3.00\%    \\
Stage 2        & 761.4              & 66.6\%       & 287820                    & 43.50\%             & 102.88            & 46.90\%   \\
Stage 3        & 88.9               & 7.8\%        & 33620                     & 5.10\%              & 8.92              & 4.10\%    \\
Control        & 208.9              & 18.3\%       & 78948                     & 11.90\%             & 19.78             & 9.00\%    \\ \hline
SRAM \\ \hline
View (12\,KB)      &                    &              & 22001                     & 3.30\%              & 7.86              & 3.60\%    \\
Process (12\,KB)   &                    &              & 22001                     & 3.30\%              & 7.86              & 3.60\%    \\
Codebook (8\,KB)  &                    &              & 15266                     & 2.30\%              & 5.43              & 2.50\%    \\
Sorting (56\,KB)   &                    &              & 113230                    & 17.10\%             & 42.74             & 19.50\%   \\
Feature (32\,KB)   &                    &              & 56759                     & 8.60\%              & 15.15             & 6.90\%    \\ \hline
\textbf{Total} & \textbf{1143.8}    & \textbf{}    & \textbf{661601}           & \textbf{}           & \textbf{219.42}   &           \\ \hline
\end{tabular}
\label{Table_areaandpower}
\end{table}

\begin{table}[btp]
\centering
\caption{Performance comparison of the 3DGS hardware implementations.
Scene set A: Tanks\&Temples and Deep Blending.
Scene set B: Set A + Mip-NeRF360 (without extra scenes like Flowers and Treehill).}
\label{Table_hardware_comparsion}
\setlength{\tabcolsep}{2pt}
\renewcommand{\arraystretch}{1.1}
\begin{tabular}{l|cc|c|c}
\hline
                         & \multicolumn{2}{c|}{Ours} & \cite{gscore} & \cite{gsnorm} \\
\cline{2-3}
\hline
Process (nm)             & \multicolumn{2}{c|}{28}                 & 28            & 22             \\
Frequency (MHz)          & \multicolumn{2}{c|}{800}                & 1000          & 500            \\
SRAM (KB)                & \multicolumn{2}{c|}{\textbf{120}}       & 272           & --             \\
Area (mm$^2$)            & \multicolumn{2}{c|}{\textbf{0.66}}      & 3.95          & 2.48           \\
Power (W)                & \multicolumn{2}{c|}{\textbf{0.219}}     & 0.87          & 0.284          \\
Resolution               & \multicolumn{2}{c|}{\textbf{1920$\times$1080}} 
                                                                   & 800$\times$600 & 800$\times$600 \\
\hline
Scene set                & A & B & A & A \\ \hline
FPS                      & \textbf{181} & \textbf{129} & 182 & 94.6 \\
Thrpt. (Mpix/s)          & \textbf{378.9} & \textbf{267.5} & 87 & 45 \\
Efficiency (Gpixels/W)   & \textbf{1.7}   & \textbf{1.2}   & 0.1 & 0.159 \\
\hline
\end{tabular}
\end{table}

\begin{table*}[tbp]
\centering
\caption{Our proposed compressed 3DGS model compared with the state-of-the-art methods.}
\label{table_our3DGS_comparision}
\begin{tabular}{l|cccc|cccc|cccc}
\hline
Dataset       & \multicolumn{4}{c|}{Mip-NeRF360}                               & \multicolumn{4}{c|}{Tanks\&Temples}                             & \multicolumn{4}{c}{Deep Blending}                               \\ \hline
Method|Metric & PSNR$\uparrow$          & SSIM$\uparrow$         & LPIPS$\downarrow$        & Size$\downarrow$         & PSNR$\uparrow$          & SSIM$\uparrow$         & LPIPS$\downarrow$        & Size$\downarrow$          & PSNR$\uparrow$          & SSIM$\uparrow$         & LPIPS$\downarrow$        & Size$\downarrow$          \\
Plenoxels         & 23.08  & 0.63  & 0.46  & 2.1\,GB  & 21.08  & 0.72  & 0.38  & 2.3\,GB  & 23.06  & 0.80  & 0.51  & 2.7\,GB \\
M-NeRF360         & 27.69  & 0.79  & 0.24  & 8.6\,MB  & 22.22  & 0.76  & 0.26  & 8.6\,MB  & 29.40  & 0.90  & 0.25  & 8.6\,MB \\
3DGS-30K          & 27.21  & 0.82  & 0.21  & 734\,MB  & 23.14  & 0.84  & 0.18  & 411\,MB  & 29.41  & 0.90  & 0.24  & 676\,MB \\
CodecGS           & 27.30  & 0.810 & 0.236 & 10.3\,MB & 23.63  & 0.841 & 0.192 & 7.8\,MB  & 29.81  & 0.906 & 0.251 & 9.0\,MB \\
ContextGS\_lowrate& 27.62  & 0.808 & 0.237 & 13.3\,MB & 24.12  & 0.849 & 0.186 & 9.9\,MB  & 30.09  & 0.907 & 0.265 & 3.7\,MB \\
HAC-highrate      & 27.77  & 0.811 & 0.230 & 22.9\,MB & 24.40  & 0.853 & 0.177 & 11.8\,MB & 30.34  & 0.906 & 0.258 & 6.7\,MB \\
EAGLES            & 27.23  & 0.81  & 0.24  & 54\,MB   & 23.37  & 0.84  & 0.20  & 29\,MB   & 29.86  & 0.91  & 0.25  & 52\,MB \\
LightGaussian     & 27.28  & 0.81  & 0.243 & 42\,MB   & 23.11  & 0.817 & 0.231 & 22\,MB   & --     & --     & --     & -- \\ \hline
Ours              & 26.247 & 0.786 & 0.20  & 14\,MB   & 23.51  & 0.83  & 0.23  & 4.5\,MB  & 28.58  & 0.89  & 0.26  & 5.0\,MB \\ \hline
\end{tabular}
\end{table*}

\begin{table}[tbp]
\centering
\caption{Ablation studies on iterative pruning, SH knowledge distillation, and VQ.}
\label{table_size_ablation}
\setlength{\tabcolsep}{2pt}
\begin{tabular}{lcccc}
\hline
\textbf{Model} & \textbf{Size$\downarrow$} & \textbf{PSNR$\uparrow$} & \textbf{SSIM$\uparrow$} & \textbf{LPIPS$\downarrow$} \\ \hline
Baseline         & (1.0$\times$)            & 26.918          & 0.832          & 0.214            \\
+Pruning         & (5.8$\times$)            & 27.144 (+0.23)  & 0.824 (-0.01)  & 0.244 (+0.03)    \\
+SH degree2      & (1.5$\times$)             & 26.974 (-0.17)  & 0.826 (+0.00)  & 0.244 (+0.00)    \\
+SH degree1      & (1.6$\times$)             & 26.668 (-0.31)  & 0.821 (-0.01)  & 0.248 (+0.00)    \\
+VQ              & (3.7$\times$)             & 26.185 (-0.48)  & 0.810 (-0.01)  & 0.259 (+0.011)    \\
Ours             & (51.6$\times$)            & 26.185 (-0.73)  & 0.810 (-0.02)  & 0.259 (+0.045)    \\ \hline
\end{tabular}
\end{table}

\begin{table}[tbp]
\centering
\caption{Impact of SH degree reduction on elements, parameters, computation, and PSNR drop.}
\begin{tabular}{|c|c|c|c|c|}
\hline
SHs Degree & Elements & Parameters & Computation & PSNR Drop \\ \hline
3 to 2     & -21     & -36\%     & -35\%       & -0.17\,dB   \\
2 to 1     & -15     & -25\%     & -17\%       & -0.31\,dB   \\ \hline
3 to 1     & -36     & -61\%     & -52\%       & -0.48\,dB   \\ \hline
\end{tabular}
\label{table_SHdegree_KD}
\end{table}

\begin{table}[htbp]
\centering
\caption{
Example analysis of iterative pruning on scene ``Stump''. This table illustrates pruning-schedule selection; multi-scene results across datasets are reported in Table~\ref{table_prune_result}.
}
\resizebox{\columnwidth}{!}{%
\begin{tabular}{|l|c|c|c|c|c|}
\hline
\textbf{Case}         & \textbf{\#GP Before} & \textbf{\#GP After} & \textbf{PSNR} & \textbf{SSIM} & \textbf{Size (MB)} \\ \hline
3DGS                  & -                    & -                   & 26.55         & 0.78          & 1068.25            \\ \hline
Prune 0.88            & 4516690             & 564127              & 25.72         & 0.70          & 133.42             \\ \hline
Iter1 0.4             & 4516690             & 2710014             & 26.83         & 0.78          & 640.95             \\
Iter2 0.4             & 2710014             & 1626008             & 26.69         & 0.78          & 384.57             \\
Iter3 0.4             & 1626008             & 975605              & 26.52         & 0.77          & 230.74             \\
Iter4 0.4             & 975605              & 565363              & 26.21         & 0.75          & 138.45             \\ \hline
Iter4 0.3             & 975605              & 682924              & 26.36         & 0.76          & 161.52             \\
Iter4 0.25            & 975605              & 731704              & 26.41         & 0.76          & 173.06             \\
\textbf{Iter4 0.2}    & \textbf{975605}      & \textbf{780484}     & \textbf{26.46} & \textbf{0.76}  & \textbf{184.59}     \\
Iter4 0.1             & 975605              & 878045              & 26.51         & 0.77          & 207.67             \\ \hline
\end{tabular}%
}
\label{table_stump_iteration_prune}
\end{table}

\begin{table*}[tbp]
\centering
\caption{Iterative pruning and fine-tuning results.}
\begin{tabular}{|c|c|cc|cc|cc|}
\hline
\textbf{Dataset} & \textbf{Scene} & \multicolumn{2}{c|}{\textbf{PSNR}} & \multicolumn{2}{c|}{\textbf{SSIM}} & \multicolumn{2}{c|}{\textbf{LPIPS}} \\ \hline
\multicolumn{1}{|c|}{} & & \multicolumn{1}{c|}{\textbf{Value}} & \textbf{Change} & \multicolumn{1}{c|}{\textbf{Value}} & \textbf{Change} & \multicolumn{1}{c|}{\textbf{Value}} & \textbf{Change} \\ \hline
\multirow{9}{*}{MipNeRF-360} 
 & Bicycle  & 25.266 & 0.02   & 0.748 & -0.023 & 0.268 & 0.063 \\
 & Bonsai   & 31.48  & -0.5   & 0.929 & -0.009 & 0.216 & 0.011 \\
 & Counter  & 28.477 & -0.223 & 0.886 & -0.019 & 0.237 & 0.033 \\
 & Garden   & 26.764 & -0.646 & 0.825 & -0.043 & 0.174 & 0.071 \\
 & Kitchen  & 30.9   & 0.583  & 0.913 & -0.009 & 0.161 & 0.032 \\
 & Room     & 31.521 & 0.889  & 0.915 & 0.001  & 0.233 & 0.013 \\
 & Stump    & 26.459 & -0.091 & 0.764 & -0.011 & 0.271 & 0.061 \\
 & Flowers  & 21.37  & -0.15  & 0.593 & -0.012 & 0.348 & 0.012 \\
 & Treehill & 22.37  & -0.12  & 0.628 & -0.010 & 0.329 & 0.012 \\ \hline
\multirow{2}{*}{Tanks\&Temples} 
 & Train & 22.692 & 1.595 & 0.807 & 0.005 & 0.262 & 0.044 \\
 & Truck & 26.202 & 1.015 & 0.892 & 0.013 & 0.163 & 0.015 \\ \hline
\multirow{2}{*}{Deep Blending} 
 & Drjohnson & 29.309 & 0.543 & 0.903 & 0.004 & 0.255 & 0.011 \\
 & Playroom  & 30.055 & 0.011 & 0.910 & 0.004 & 0.249 & 0.008 \\ \hline
\multicolumn{2}{|c|}{Average} 
 & 27.143 & 0.225 & 0.824 & -0.008 & 0.244 & 0.030 \\ \hline
\end{tabular}
\label{table_prune_result}
\end{table*}

\begin{table*}[tbp]
\centering
\caption{Performance and compression results with the proposed quantization.}
\begin{tabular}{|cc|c|c|c|}
\hline
\multicolumn{1}{|c|}{\textbf{Dataset}} & \textbf{Scene} & \textbf{PSNR} & \textbf{SSIM} & \textbf{LPIPS} \\ \hline
\multicolumn{1}{|c|}{\multirow{9}{*}{MipNeRF-360}}   & Bicycle        & 24.793           & 0.737            & 0.284             \\
\multicolumn{1}{|c|}{}                               & Bonsai         & 29.647           & 0.913            & 0.228             \\
\multicolumn{1}{|c|}{}                               & Counter        & 27.034           & 0.862            & 0.253             \\
\multicolumn{1}{|c|}{}                               & Garden         & 26.163           & 0.811            & 0.194             \\
\multicolumn{1}{|c|}{}                               & Kitchen        & 29.357           & 0.896            & 0.172             \\
\multicolumn{1}{|c|}{}                               & Room           & 30.387           & 0.896            & 0.241             \\
\multicolumn{1}{|c|}{}                               & Stump          & 26.082           & 0.756            & 0.294             \\ 
\multicolumn{1}{|c|}{}                               & Flowers        & 20.881           & 0.582            & 0.369             \\ 
\multicolumn{1}{|c|}{}                               & Treehill       & 21.881           & 0.617            & 0.350             \\ \hline
\multicolumn{1}{|c|}{\multirow{2}{*}{Tanks\&Temples}}& Train          & 21.672           & 0.791            & 0.279             \\
\multicolumn{1}{|c|}{}                               & Truck          & 25.358           & 0.878            & 0.181             \\ \hline
\multicolumn{1}{|c|}{\multirow{2}{*}{Deep Blending}} & Drjohnson      & 28.217           & 0.894            & 0.266             \\
\multicolumn{1}{|c|}{}                               & Playroom       & 28.937           & 0.895            & 0.254             \\ \hline
\multicolumn{2}{|c|}{Average}                        & \textbf{26.185}  & \textbf{0.810}   & \textbf{0.259}    \\ \hline
\end{tabular}
\label{table_vq_result}
\end{table*}

\begin{figure*}[tbp]
    \centering
    \begin{minipage}{0.48\linewidth}
        \centering
        \includegraphics[width=\linewidth]{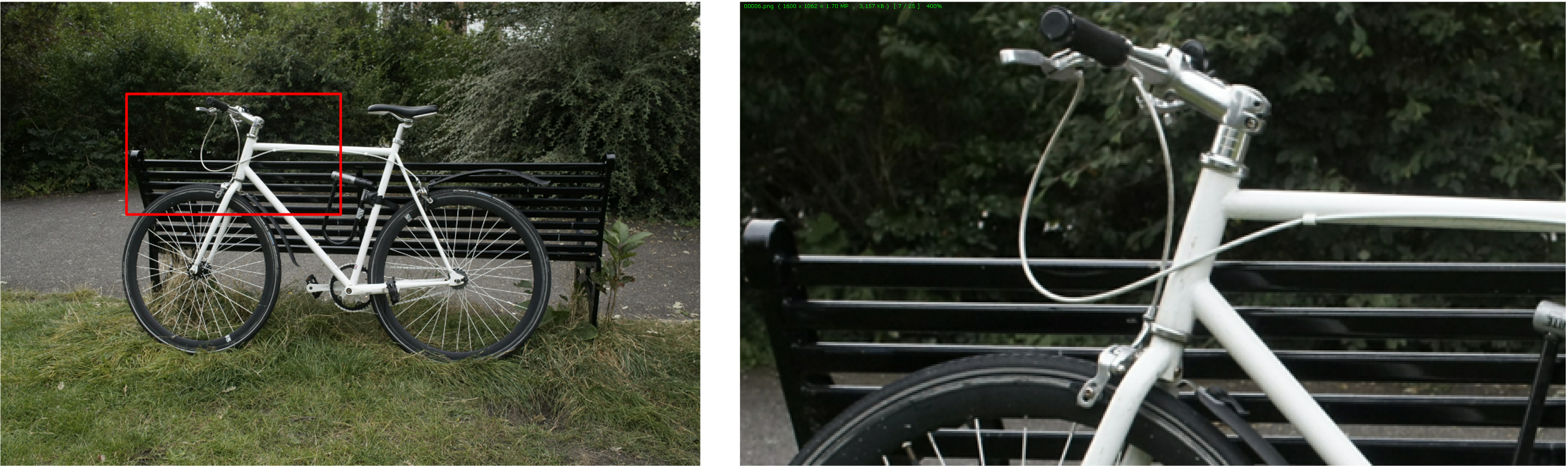}
        \label{vq_bicycle_GT}
    \end{minipage}
    \hspace{0.02\linewidth}
    \begin{minipage}{0.48\linewidth}
        \centering
        \includegraphics[width=\linewidth]{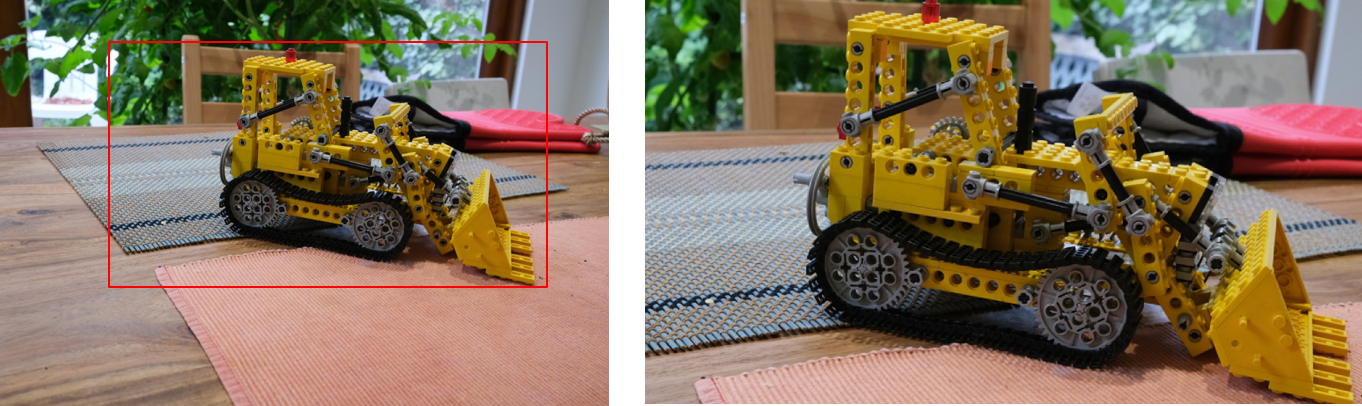}
        \label{vq_kitchen_gt}
    \end{minipage}
    
    \begin{minipage}{0.48\linewidth}
        \centering
        \includegraphics[width=\linewidth]{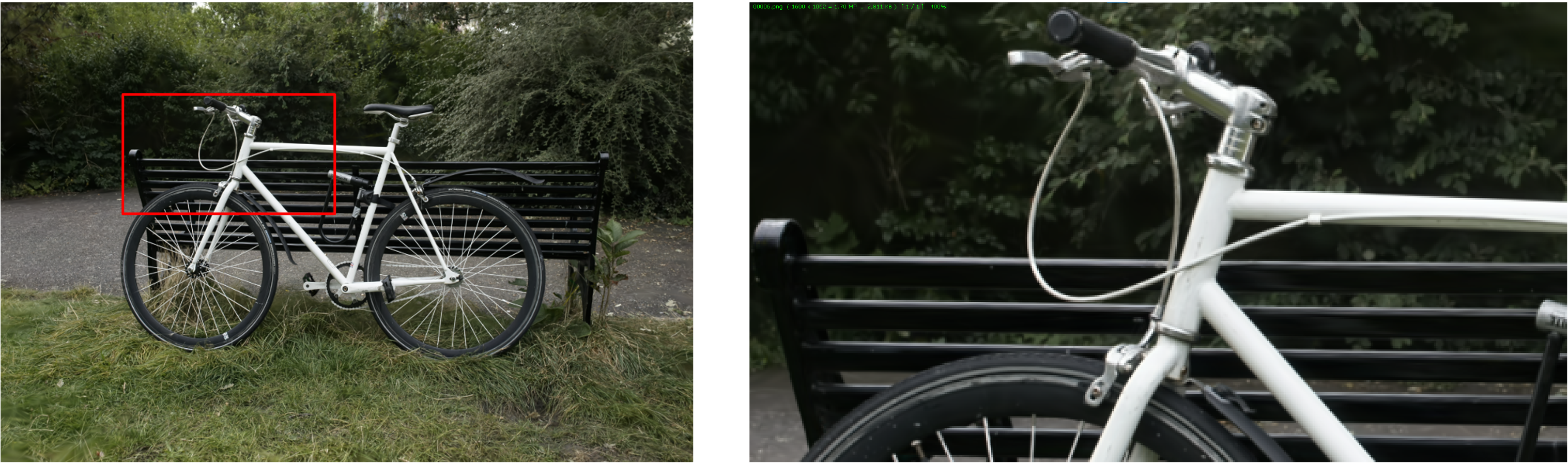}
        \label{vq_bicycle_3dgs}
    \end{minipage}
    \hspace{0.02\linewidth}
    \begin{minipage}{0.48\linewidth}
        \centering
        \includegraphics[width=\linewidth]{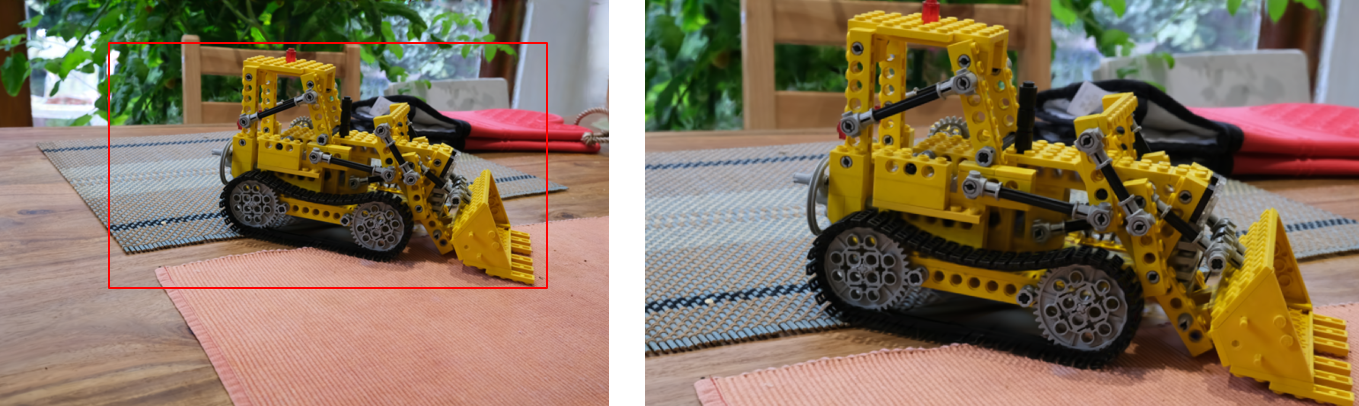}
        \label{vq_kitchen_3dgs}
    \end{minipage}
    
    \begin{minipage}{0.48\linewidth}
        \centering
        \includegraphics[width=\linewidth]{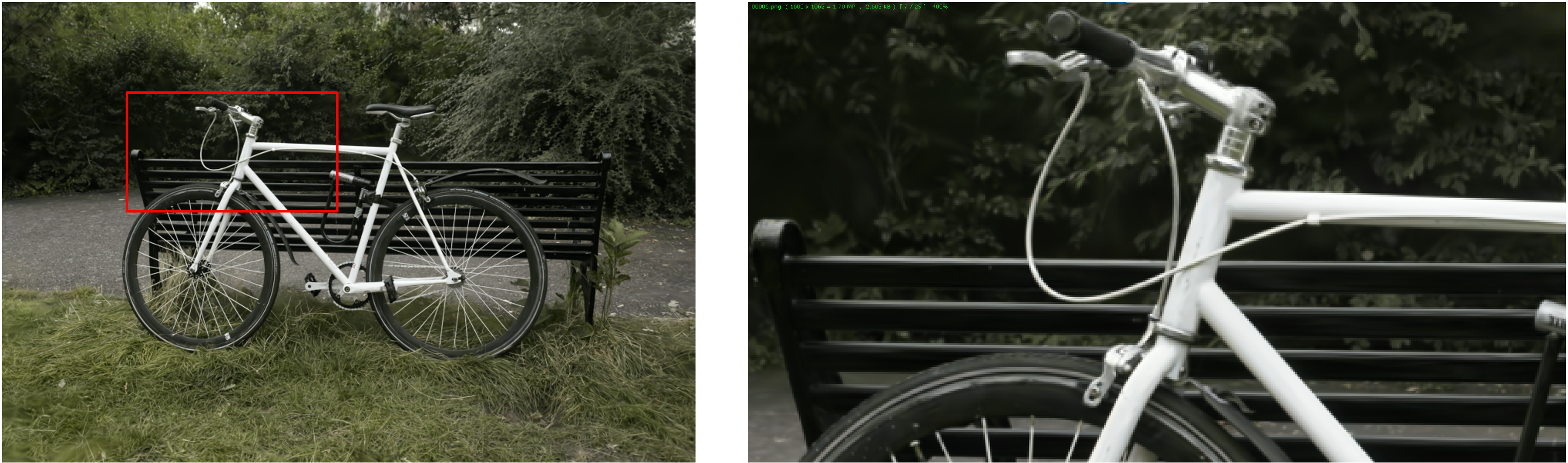}
        \label{vq_bicycle_ours}
    \end{minipage}
    \hspace{0.02\linewidth}
    \begin{minipage}{0.48\linewidth}
        \centering
        \includegraphics[width=\linewidth]{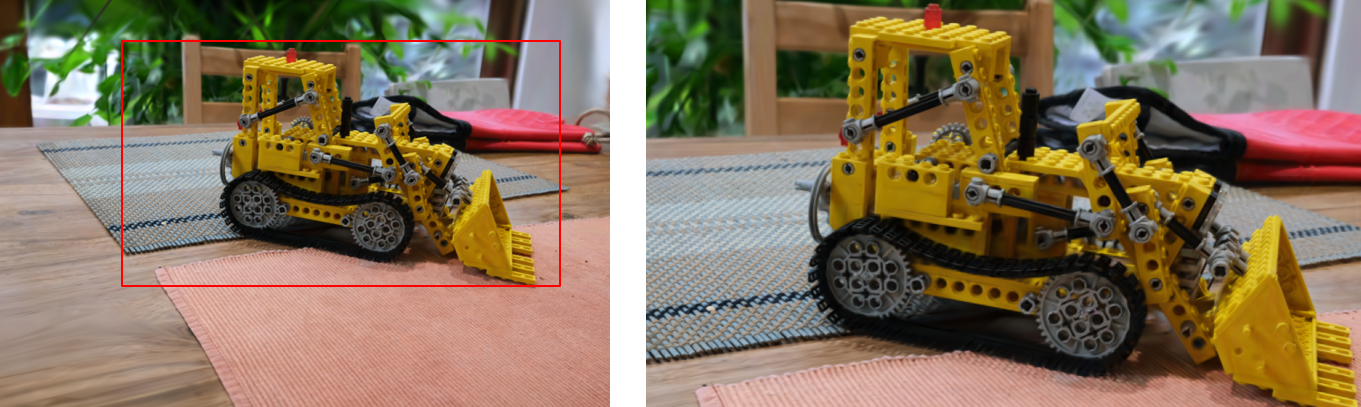}
        \label{vq_kitchen_ours}
    \end{minipage}
    
    \caption{Visual comparison of the results. Top to bottom: ground truth, 3DGS~\cite{3DGS}, and our compressed 3DGS model for the Bicycle (left) and Kitchen (right) scenes.}
    \label{fig:vq_comparison}
\end{figure*}

\subsection{Hardware Implementation Results}
\label{5-subsection:Hardware Implementation results}
We implemented the design in Verilog and synthesized it with the Synopsys Design Compiler using TSMC-28\,nm CMOS technology at 800\,MHz.  Our proposed design is intended primarily for full-HD (FHD) resolution, with a tile size of 16$\times$16 pixels. Therefore, each frame consists of 120$\times$68 = 8160 tiles. Given that both steps are optimized through a pipelined architecture, our proposed hardware design is capable of achieving a frame rate of 129\,FPS at the FHD resolution. Note that we do not have a fabricated silicon prototype for measurement due to cost reason. All the hardware related numbers are based on the gate level simulation.

\subsubsection{Area and Power Analysis}
Table~\ref{Table_areaandpower} presents the synthesis results for gate count, area, and power consumption. The design uses 1143.8k gate with 120\,KB SRAM. The table shows that Stage 2 
now stands out as the most resource-intensive part of the architecture after our optimizations,   
both in terms of area and power consumption. It accounts for approximately 43.5\% of the total area consumption and 46.9\% of the total power consumption. Despite Stage 2 being the bottleneck of the overall flow, the decision to use the 4$\times$ parallelism in Stages 2 and 3 helps to strike a balance between speed and power efficiency. This approach results in a total power consumption of 219.41\,mW and an area of 0.66\,mm$^2$ throughout the architecture, ensuring that the resource requirements of speed, power, and area are aligned with the expected design goals.

\subsubsection{Comparison to Other Works}
\label{5-subsection:Design comparison}
Table~\ref{Table_hardware_comparsion} shows a comparison with other hardware designs. The frame rate and throughput are average values based on the dataset. 
 Our design achieves 4.355$\times$ higher throughput than~\cite{gscore} and 10.69$\times$ higher energy efficiency than~\cite{gsnorm} for the same scene case  while requiring 3.75$\times$ smaller area than~\cite{gsnorm}. In comparison, while other designs~\cite{gscore, gsnorm} have tried to reduce computation by hierarchical sorting~\cite{gscore} and splat normalization~\cite{gscore}, their optimizations are limited and thus still require large area and power. Reference~\cite{LeeGS} is not included in this table due to its FPGA implementation. 
Our advantages lie not only in the compression of the model, but also in the optimization of dataflow and individual units, which results in a highly efficient design for edge devices.

\subsection{Ablation Study}
\subsubsection{Compressed 3D Gaussian Splatting Model Results}
\label{5-subsection:Compressed 3D Gaussian Splatting Model results}

Table~\ref{table_size_ablation} reports the average performance of our proposed compressed 3DGS model on the evaluated datasets. The baseline model (3DGS-30K) refers to the original 3DGS model~\cite{3DGS} evaluated on the Mip-NeRF360, Tanks\&Temples, and Deep Blending datasets. As shown in the table, iterative pruning significantly reduces model size by 5.8$\times$. Applying SH knowledge distillation at different degrees further reduces model size with a gradual quality--compression tradeoff, and VQ provides an additional 3.7$\times$ compression. In particular, increasing compression from the LightGaussian-level regime (13.92$\times$) to our final 51.6$\times$ operating point yields an additional 3.7$\times$ size reduction while incurring only a 0.48\,dB PSNR drop on average. Overall, our proposed compressed model is 51.6$\times$ smaller than the baseline, with a 0.743\,dB drop in PSNR, while maintaining competitive performance across different quality metrics.  These results show that the remaining compression beyond the LightGaussian-level regime provides a substantial reduction in representation size at a bounded quality cost, which is favorable for our targeted edge hardware setting.

Table~\ref{table_stump_iteration_prune} presents an illustrative example of the iterative pruning and fine-tuning schedule on the ``Stump'' scene. Iter1 refers to the first pruning/fine-tuning iteration, in which a pruning rate of 0.4 removes 40\% of the Gaussian points with only a minor PSNR drop. Iter2 and Iter3 follow the same procedure. In Iter4, we further evaluated several candidate pruning rates, including 0.4, 0.3, 0.25, 0.2, and 0.1. Considering the tradeoff between point removal and quality loss, we selected a pruning rate of 0.2 for Iter4. This example illustrates how the final four-step pruning schedule was determined.

The main evidence for the robustness of the pruning strategy is provided by the multi-scene results across datasets.  Detailed changes in PSNR, SSIM, and LPIPS for each scene are reported in Table~\ref{table_prune_result} for iterative pruning and fine-tuning, Table~\ref{table_SHdegree_KD} for SH knowledge distillation, and Table~\ref{table_vq_result} for VQ. As shown in Table~\ref{table_prune_result}, the proposed four-step iterative pruning process removes 87\% of Gaussian points and reduces model size by 82.7\% while maintaining, and in some cases improving, image quality relative to the original 3DGS results. After pruning, SH knowledge distillation in Table~\ref{table_SHdegree_KD} further reduces 61\% of the parameters and 52\% of the computation with a 0.48\,dB PSNR drop. Finally, VQ in Table~\ref{table_vq_result} further reduces model size by 73.04\% with an additional 0.48\,dB PSNR drop. The resulting size-reduction and Gaussian-removal ratios are fixed across scenes, which simplifies the hardware design and helps maintain a predictable low-power implementation target.

\subsubsection{Visual Results}
\label{5-section:Experimental Results}

Fig.~\ref{fig:vq_comparison} shows the visual results of the proposed compression method. Compared with GT and 3DGS, our reconstruction largely preserves the global appearance of the rendered scenes, including overall color balance, brightness, and contrast at the full-frame level. The remaining differences are mainly concentrated in high-frequency details and local textures: very fine structures appear slightly smoother and less crisp, with a small reduction in micro-contrast. In particular, around the metallic handlebar, our result exhibits weaker and more blurred specular highlights than those in GT and 3DGS, making the surface appear slightly less shiny and more matte. In the grass region, our compressed model shows a slightly warmer, more yellowish-green tone than GT and 3DGS, indicating a minor color shift due to compression. Overall, these differences are subtle and become noticeable mainly under close side-by-side inspection, while at normal viewing distance the renderings remain visually very similar.

\subsubsection{Performance comparison}
\label{sec_performance_comparison}

Table~\ref{table_our3DGS_comparision} compares our proposed compressed 3DGS model with recent compression methods using PSNR, SSIM, learned perceptual image patch similarity (LPIPS), and model size. The results show that different methods occupy different quality--size operating points. Although our image quality is not the highest, our method is particularly effective in reducing model size. Compared with 3DGS~\cite{3DGS}, we reduce the model size substantially, with an average PSNR drop of 0.743\,dB across the three evaluated datasets. Compared with LightGaussian~\cite{lightgaussian}, we achieve a substantially higher compression ratio (51.6$\times$ vs.\ 15$\times$) with an average PSNR drop of 0.832\,dB on Mip-NeRF360 and Tanks\&Temples. These results indicate that our method occupies a hardware-oriented operating point, favoring a much smaller representation at a bounded quality cost for low-power edge deployment.

\subsubsection{Why Aggressive Compression Is Necessary for Real-Time Low-Power Edge Deployment}
\label{sec:why_compression_power}

We further justify the necessity of compression from a low-power edge perspective. At real-time frame rates, energy is driven by both per-Gaussian-point processing and the movement of Gaussian parameters through the pipeline. Table~\ref{tab:power_edge_feasibility} summarizes the result. Even a LightGaussian-level operating point would still imply a 3.71$\times$ increase in work relative to our final design, corresponding to about 0.81~W at 1080p@129~FPS. This is substantially higher than our 0.219~W result and no longer compatible with the intended low-power edge setting. Equivalently, if the same power regime were enforced, throughput would have to decrease proportionally. Therefore, additional compression beyond the LightGaussian-level regime is necessary not only to reduce model size, but also to sustain real-time performance under the targeted edge power budget.

\begin{table}[t]
\centering
\caption{Power analysis of compression for edge deployment ith the same FPS at comparable efficiency. 
}
\label{tab:power_edge_feasibility}
\begin{tabular}{l c c c}
\toprule
Configuration & Ratio ($\times$) & Implied power (W)  \\
\midrule
Ours (pruning + SH + VQ) & 1.00 & 0.219 \\
LightGaussian & 3.71 & 0.812  \\
w/o pruning (SH + VQ only) & 7.69 & 1.68  \\
w/o SH+VQ (pruning only) & 6.71 & 1.47  \\
Uncompressed & 51.6 & 11.3  \\
\bottomrule
\end{tabular}
\end{table}

\subsubsection{Hardware Component Ablation}
\label{sec:hw_component_ablation}

Table~\ref{tab:hw_ablation_sequential} presents a hardware ablation analysis normalized to the final reported throughput. Starting from the baseline with none of the three optimizations enabled, we add near-plane culling, zero-Jacobian skipping, and early termination in pipeline order. The incremental gain column shows the contribution of each step relative to the previous configuration. Near-plane culling provides the largest early gain by reducing the number of Gaussian points entering the pipeline, zero-Jacobian skipping substantially improves projection-side efficiency, and early termination provides the final throughput gain by eliminating unnecessary blending work. Under this first-order model, the three optimizations together explain the progression from 20.4 FPS to the final 129 FPS design point.

\begin{table}[t]
\centering
\caption{Hardware ablation analysis normalized to the final reported throughput. 
}
\label{tab:hw_ablation_sequential}
\begin{tabular}{l c c c}
\toprule
Configuration & throughput (\%) & FPS@1080p & gain ($\times$) \\
\midrule
Baseline (none) & 15.81 & 20.4 & -- \\
+ culling & 35.94 & 46.4 & 2.27 \\
+ zero-Jacobian & 75.70 & 97.7 & 2.11 \\
+ early term.(full) & 100.00 & 129.0 & 1.32 \\
\bottomrule
\end{tabular}
\end{table}

\section{Conclusion}
\label{chapter:conclusion}

We presented a hardware accelerator for 3DGS to enable efficient real-time rendering. Our proposed compression techniques achieved a \(51.6\times\) model-size reduction, resulting in a compact representation with a 0.743~dB PSNR drop, effectively balancing memory efficiency and visual quality. The dedicated hardware architecture significantly improved performance over existing designs with frame-level pipeline, mixed granularity processing, and avoidance of unnecessary computation and memory access, delivering a \(3.14\times\) higher throughput, \(7.5\times\) better energy efficiency and \(3.75\times\) smaller silicon area. Implemented in a TSMC 28~nm process at 800~MHz, the accelerator achieved 267.5~Mpixels/s throughput, 0.219~W power consumption, and an energy efficiency of 1219~Mpixels/J, supporting full-HD rendering at 129~FPS within an area of 0.66~mm\textsuperscript{2}.

\bibliographystyle{IEEEtran}

\bibliography{bib/ieeeBSTcontrol,bib/thesis}



	
	

\end{document}